\documentclass[review,authoryear]{elsarticle}  
\usepackage[latin1]{inputenc}
\usepackage[english]{babel}
\usepackage{geometry}
\usepackage{subfigure}
\usepackage{textcomp}
\usepackage{pifont}
\usepackage[leqno]{amsmath}
\usepackage{dsfont}
\usepackage{amstext}
\usepackage{amsfonts}
\usepackage{latexsym}
\usepackage{graphicx}

\usepackage{url}
\usepackage{float}
\usepackage{subfigure}
\usepackage{amssymb}
\usepackage{fancyvrb}

\graphicspath{{./images/}}
\title{Estimating the Algorithmic Complexity of Stock Markets}

\author{
Olivier~Brandouy $^\dag$,
        Jean-Paul Delahaye$^\ddag$,
        Lin Ma$^\ast$
\\\vspace{12pt}  \normalfont{$^\dag$ University of Bordeaux 4, \url{olivier.brandouy@u-bordeaux.fr}\\
\normalfont{$^\ddag$} University of Lille 1, \url{delahaye@lifl.fr}\\
$^\ast$ Corresponding author, University of Lille 1, \url{lin.ma@iae.univ-lille1.fr}
}
}
\begin{document}

 \begin{abstract}
Randomness and regularities in Finance are usually treated in
probabilistic terms. In this paper, we develop a completely
different approach in using a non-probabilistic framework based on
the algorithmic information theory initially developed  by
\citet{Kolmo1965}. We present some elements of this theory and show
why it is particularly relevant to Finance, and potentially to other
sub-fields of Economics as well. We develop a generic method to
estimate the Kolmogorov complexity of numeric series. This approach
is based on an iterative ``regularity erasing procedure''
implemented to use lossless compression algorithms on financial
data. Examples are provided with both simulated and real-world
financial time series. The contributions of this article are
twofold. The first one is methodological : we show that some
structural regularities, invisible with classical statistical tests,
can be detected by this algorithmic method. The second one consists
in illustrations on the daily Dow-Jones Index suggesting that beyond
several well-known regularities, hidden structure may in this index
remain to be identified.
 \end{abstract}


\maketitle \pagebreak

 \vspace{0.5cm}
\section*{Introduction}
Consider the following strings consisting of  $0$ and $1$:

\begin{center}
\begin{description}\label{int}
\item[A:] \texttt{01010101010101010101010101010101010101010101010101010101010101010101010101010101}
\item[B:] \texttt{01101001100101101001011001101001100101100110100101101001100101101001011001101001}
\item[C:] \texttt{01101110010111011110001001101010111100110111101111100001000110010100111010010101}
\item[D:] \texttt{11001001000011111101101010100010001000010110100011000010001101001100010011000110}
\item[E:] \texttt{00111110010011010100000000000111001111111001110010011000001001001001011101011011}
\item[F:] \texttt{00100001100000000110010010101000101110111101000011111001001000110111011110010100}
\end{description}
\end{center}
Some of these sequences are generated by simple mathematical
procedures, others are not. Some exhibit obvious regularities, and
others might reveal ``structures'' only after complicated
transformations. Only one of them is generated by a series of random
draws. The question is ``how to identify this latter'' ?

In this paper, a generic methodology will be introduced to tackle
this question of distinguishing regular (structured, organized)
sequences from random ones. Although this method could have a
general use, here, it is geared to regularity detection in Finance.
To illustrate this idea, examples are provided using both simulated
and real-world financial data. It is shown that some structural
regularities, undetectable by classical statistical tests, can be
revealed by this new approach.

Our method is based on the seminal works of Andrei \citet{Kolmo1965}
who proposed a definition of random sequences in non-statistical
terms. His definition, which is actually among the most general
ones, is itself based on Turing's and Godel's contributions to the
so-called ``computability theory'' or ``recursion theory''
\citep{turing36,godel31}. Kolmogorov's ideas have been exploited by
physicists for a renewed definition of entropy \citep{Zurek1989}, by
biologists for the classification of phylogenic trees
\citep{cilibra2005} and by psychologists to estimate
randomness-controlling difficulties \citep{griften2001}.

In Finance, Kolmogrov complexity is often used to measure the
possibility to predict future returns. For example, \citet{Chen1996}
or \citet{Chen1999} estimated the stochastic complexity\footnote{The
notion of stochastic complexity is proposed by \citet{Rissanen1986}
in replacing the universal Turing machine in the definition of
Kolmogorov complexity by a class of \emph{probabilistic} models.} of
a stock market using the sum of the squared prediction errors
obtained by econometric models. \citet{Azhar1994} measured the
complexity of stock markets with the highest successful prediction
rate\footnote{To obtain this successful prediction rate, at each
step, the author uses compression algorithms to predict the
direction of the next return, and calculates the rate of successful
predictions for the whole series.} (SPR) that one can achieve with
different compression techniques. \citet{Shmilovici2003} and
\citet{Shmilovici2009} used the Variable Order Markov model
(\emph{VOM}, a variant of context predicting compression tools) to
predict the direction of financial returns. They found a significant
difference between the SPR obtained from financial data and that
obtained from random strings. To establish a formal link between
this result and the Efficient Market Hypothesis (EMH), the authors
also simulated VOM-based trading rules on Forex time series and
concluded that there were no abnormal profits.

Exploiting another compression technique, \citet{Silva2008} and
\citet{Giglio2008} ranked stock markets all over the world according
to their LZ index, an indicator showing how well the compression
algorithm proposed by \citet{Lempel_Ziv_1976} works on
 financial returns.

Despite the perspectives opened up by these pioneer works,
two main limits in the aforementioned literature can be
highlighted:
\begin{enumerate}
    \item From a theoretical point of view, the frontier between the
 probabilistic framework and the algorithmic one is not clearly
 established.

 A strength of the algorithmic complexity is that it tackles \emph{one given string} at a time, and not necessarily a
 \emph{population of strings} with probabilities generated by a given stochastic
 process. Hence, no probabilistic assumption is needed when one uses
algorithmic complexity.

 The estimation of successful prediction rates seems to suggest that
 price motions follow a certain distribution law. Despite the use of
 compression tools, this technique reintroduces a probabilistic framework. The general and non-probabilistic framework
proposed by the algorithmic complexity theory is then weakened.

 \item This is not the case with \citet{Silva2008} and \citet{Giglio2008}.
 However, the discretization technique used in these papers
 remains open to discussion. Actually, financial returns are often
 expressed in real numbers\footnote{For example, continuous returns are computed as the logarithm of the ratio between consecutive prices :
$r_t=\log(p_{t+1})-\log(p_{t})$.}, while
compression tools only
 deal with integer numbers. So, regardless the use of compression tools, a discretization process which transforms real-number series into discrete ones, is always necessary.

 To fit this requirement, \citet{Shmilovici2003} or \citet{Shmilovici2009}, as well as
   \citet{Silva2008} or \citet{Giglio2008}, proposed to transform financial
 returns into 3 signals: ``positive", ``negative", or ``stable"
 returns.

 Undoubtedly, this radical change leads to a significant loss of information from the original financial series. As \citet{Shmilovici2009} remarked themselves, \emph{``the main
limitation of the VOM model is that it ignores the actual value of
 the expected returns
\citep{Shmilovici2009}."}

The introduction of algorithmic complexity in finance could have
wider implications. For example,
\citet{Dionisio_Menezes_Mendes_2007} have claimed that the notion of
complexity could become a measure of financial risk - as an
alternative to ``value at risk" or ``standard deviation" - which
could have general implications for portfolio management (see
\citet{Groth2011} for an application in this sense).
\end{enumerate}

Given these two points, it seems interesting to establish a general
algorithmic framework for price motions. We propose an empirical
method that allows to treat financial time series with algorithmic
tools, avoiding the over-discretisation problem. This method is
based on the initial investigation of \citet{brandouy2005} whose
idea is developed by \citet{ma2010} on tackling real-world financial
data. In this same framework, \citet{Zenil2011} fostered future
applications of Kolmogorov complexity in Finance.

Using the algorithmic approach, we show that daily returns of Dow
Jones industrial index has a relatively high Komogorov complexity,
especially after erasing the most documented stylized facts
\citet{Rama2001}. This result supports the Efficient Market
Hypothesis \citet{Fama1970} which attests the impossibility to
outperform the market.

This paper is organized as follows:

A first section formally presents Kolmogorov complexity and provides
an elementary illustration of this concept. A second section
proposes a generic methodology to estimate Kolmogorov complexity of
numeric series. We then show, in a third section, that some
regularities, difficult to detect with traditional statistic tests,
can be identified by compression algorithms. The last section is
dedicated to the interests and limits of our techniques for real
financial data analysis.
\begin{center}

\fbox{ 
\begin{minipage}[c]{14cm}
Before our formal presentation of Kolmogorov complexity, readers might be interested by the solution to our initial question: ``how to identify, among sequences $A$, $B$, $C$, $D$, $E$ and $F$, the randomly generated one?''\\

The first four sequences (or strings) are \emph{simple} in the sense of Kolmogorov, since
they can be described by the following rules:
\begin{center}
    \begin{itemize}
        \item obviously, $A$ is a 40-time repetition of \texttt{"01"} .
        \item $B$ is the so-called ``Thue-Morse'' sequence
        (see \citet{allouche2000}), which is iteratively generated as follows:
            \begin{enumerate}
                \item the initial digit is 0.
                \item the negation of ``0'' (\emph{resp.} ``1'') is ``1'' (\emph{resp.} ``0'') .
                \item to generate $B$ in a recursive way, take the existing digits, and add their negations on the
                right:
                    \begin{itemize}
                        \item Step 1: ``0''$\leftarrow$``1''
                        $\Rightarrow$ ``01''
                        \item Step 2: ``01''$\leftarrow$``10'' $\Rightarrow$ ``0110''
                        \item Step 3: ``0110''$\leftarrow$``1001'' $\Rightarrow$ ``01101001''
                        \item Step
                        4: ``01101001''$\leftarrow$``10010110'' $\Rightarrow$ ``0110100110010110''
                        \item ...
                    \end{itemize}
            \end{enumerate}
        \item $C$ is the concatenation of natural numbers written in base 2: \texttt{0, 1, 10, 11, 100, 101, 110, 11
1000...}. It is known as the ``\citet{champernowne1933} constant'' .
        \item $D$ is generated by transforming the decimals of $\pi$
into binary digits.
    \end{itemize}
\end{center}
These generating rules make $A$, $B$, $C$ and $D$ compressible by
the corresponding algorithms. For $D$, common compression tools may
turn out to be inefficient, but it is easy to write a
$\pi$-compressing algorithm, in exploiting the generating
function(s) of the famous constant.

$E$ has been drawn from an uniform law which delivers, with a
probability of $50\%$ for each, $1$ or $0$. The last sequence ($F$)
corresponds to the daily variations of the Dow Jones industrial
index observed at closing hours from 07/10/1987 to 10/27/1987, with
\texttt{"0"} coding the drops and \texttt{"1"} the rises.

One can question here to which extent $F$ is similar to the first
four strings, and to which degree lossless compression algorithms
can distinguish them. This question serves as a baseline in this
paper which attempts to provide a new consideration of randomness in
financial dynamics.

\end{minipage}
}
\end{center}
\section{Regularity, randomness and Kolmogorov complexity}

We start this section with traditional interpretations of randomness
in Finance in order to contrast them with their counterpart in
computability theory. We then explicitly present the notion of
``Kolmogorov complexity'' and illustrate its empirical applications
with simulated data.

 \subsection{Probabilistic \emph{versus} algorithmic randomness} \label{finance}
Traditionally, financial price motions were modeled (and are still
modeled) as random walks\footnote{These processes correspond to the
so-called ``Brownian motions'' in continuous contexts. In this
paper, as real-world financial data are always discrete and
algorithmic tools only handle discrete sequences, our theoretical
developments will lie in a discrete framework.} (see equation
\ref{basic}):
\begin{equation}
\label{basic} p_t=p_{t-1}+\varepsilon_{t}
\end{equation}
In this equation, $p_t$ stands for the price of one specific
security at ``$t$'' and $\varepsilon_{t}$ is a stochastic term drawn
from a certain distribution law\footnote{Up to now, there is no
consensus in the scientific community of Finance concerning the
driving law behind $\varepsilon_{t}$ series.}. This stochastic term
is often estimated by a Gaussian noise, though it fails to correctly
represent some stylized facts such as the time dependence in
absolute returns or the auto-correlation in volatility (see for
example \citet{mandelbrot1968} or \citet{Fama1965}). Several
stylized facts are taken into account by volatility models (see, for
instance GARCH models\footnote{Generalized Auto Regressive
Conditional Heteroscedasticity Models, \citet{bollerslev1986}}),
while others stay rarely exploited in financial engineering (for
example, multi-scaling laws \citet{calvet2002}, return seasonality
$\cdots$).

In this sense, financial engineering proposes an iterative process
which, step by step, improves the quality of financial time series
models, both in terms of describing and forecasting\footnote{It is
clear that ``description'' is the easier part of the work, with
forecasting being far more complicated.}. At least from a
theoretical point of view, this iterative process should deliver, at
the end, a perfectly deterministic model\footnote{Which is simply
impossible for the moment, because of our imperfect comprehension of
financial time series.}. If this final stage symbolizes the perfect
comprehension of price dynamics, each step in this direction can be
considered as a progress of our knowledge on financial time series.
With an imperfect understanding of this latter, the iterative
process will stop somewhere before the determinist model. Meanwhile,
wherever we stop in this process, the resulting statistical model is
a ``compressed expression'' (some symbols and parameters) of the
long financial time series (hundreds or thousands of observations).

``To understand is to compress'' is actually the main idea in
Kolmogorov complexity \citep{Kolmo1965} whose definition is formally introduced as
follows:
\newtheorem{kolmogorov}{\textbf{Definition}}[section]
\begin{kolmogorov}\label{definition}
Let $s$ denote a finite binary string consisting of digits from
$\{0, 1\}$. Its Kolmogorov complexity $K(s)$ is the length of the
shortest program $P$ generating (or printing) $s$. Here, the length
of $P$ depends on its binary expression in a so-called ``universal
language''\footnote{A universal language is a programming language
in which one can code all ``calculable functions''.
``Calculability'' is often related to the notion of
``effectiveness'' in computer science (see for example
\citet{Velupillai2004}). Most modern programming languages are
universal (for example C, Java, R, Lisp $\cdots$).}.
\end{kolmogorov}
As a general indicator of randomness, Kolmogorov complexity of a
given string should be relatively stable to the choice of universal
language. This stability is shown in the following theorem (see \cite{Kolmo1965} or \cite{Li1997} for a complete proof)
according to which, a change in the universal language has always a
bounded impact on a string's Kolmogorov complexity.
\newtheorem{language1}[kolmogorov]{\textbf{Theorem}}
\begin{language1}\label{invariance}
Let $K_{L_1}(s)$ and $K_{L_2}(s)$ denote the Kolmogorov complexities
of a string $s$ respectively measured in two different universal
languages $L_{1}$ and $L_{2}$.  The difference between $K_{L_1}(s)$
and $K_{L_2}(s)$ is always a constant $c$ which is completely
independent of $s$:
\begin{equation}
\label{k1} \exists c \forall s \textrm{ }|K_{L_1}(s)-K_{L_2}(s)|\leq
c
\end{equation}
\end{language1}
According to this invariance theorem, one need not pay too much
attention to the choice of universal language in Kolmogorov
complexity estimations, since its intrinsic value (the part
depending on $s$) remains stable with regard to programming technics
in use.

Besides the invariance theorem, another property of Kolmogorov
complexity, firstly proved in \cite{Kolmo1965}, makes it a good randomness measure: a finite string's
Kolmogorov complexity always takes a finite value.

\newtheorem{maximumK}[kolmogorov]{\textbf{Theorem}}
\begin{maximumK}
If $s$ is a sequence of length $n$ then:
\begin{equation}\label{uuu}
K(s)\leq n+ O(log(n))
\end{equation}
\end{maximumK}
In equation \ref{uuu}, $O(log(n))$ only depends on the universal
language in use. This property comes from the fact that all finite
strings can at least be generated by the program, ``\texttt{print
(s)}'', whose length is close to the size of $s$.

Remark that most $n$-digit strings (provided that $n$ is big enough)
have their Kolmogorov complexity $K(s)$ close to $n$. $K(s)<<n$
implies the existence of a program $P$ which can generate $s$ and is
much shorter than $s$. In this case, $P$ is said to have found a
structural regularity in $s$, or $P$ has compressed $s$.

On the contrary,
\newtheorem{definition}[kolmogorov]{\textbf{Definition}}
\begin{definition}\label{definitionkolmogorov}
if for an \textbf{infinite binary string} denoted by $S$:
$$\exists C \forall n  \textrm{ } K(S\upharpoonright
n)\geq n - C$$ where $S\upharpoonright n$ denotes the first $n$
digits of $S$ and $C$ a constant in the natural number set $\mathds{N}$, then $S$ is
defined to be a ``random string''.
\end{definition}
This definition of random strings, 
proposed by \citet{malof66}, reformulated by \citet{Chaitin1987} and reviewed in \citet{Downey2010}, is irrelevant to any probability notion, and only depends on structural properties of $S$.

Following definition \ref{definitionkolmogorov}, the best compression rate one can get
from the $n$ first digits of a random string is calculated by
$$rate=\frac{n-K(S\upharpoonright
n)}{n}=\frac{C}{n}$$

As $n\rightarrow\infty$, $rate\rightarrow0$. In other terms, random
strings are incompressible ones.

Contra-intuitively, compressible strings are relatively seldom. For
example, among all $n$-digit strings, the proportion of those
verifying $K(s)<n-10$ does not exceed $1/1000$\footnote{The
probability to observe $K(s)<n-20$ is less than $1/1000000$. Proof
for this results are provided in appendix \ref{annexe1}.}.

To illustrate the relation between compressibility and regularity,
let's consider two compressible (regular) strings:
\begin{description}
\item[a:] \texttt{01101110010111011110001001101010111100110111101111100001000110010100111010010101}
\item[b:] \texttt{11111001111001100111101110011101101010101110111110111101101111111111101001101110}
\end{description}
The first string, known as Champernowne sequence (i.e. string $C$ on
page \pageref{int}), is generated with a quite simple arithmetic
rule: it is the juxtaposition of natural numbers written in base 2.
The optimal compression rate on the first million digits of \url{a}
is above $99\%$ (c.f. annex \ref{annexe3}), which indicates that
\url{a} can be printed by a short program.

The second string was independently drawn from a random law (with
the help of a physical procedure) delivering $``1''$ with a $2/3$
probability and $``0''$ with a $1/3$ probability. The computer
program generating \url{b} can be much shorter than it. In fact,
classical tools can obtain on \url{b} a compression rate calculated
as follows\footnote{This compression rate is related to the
Shannon's entropy of \url{b}.}:
\begin{equation}
-1/3log_2(1/3)-2/3log_2(2/3)=91,8\%
\end{equation}
While \url{b} is drawn from a random law, its compressibility
remains perfectly compatible with statistical conjectures: composed
by definitively more $1$ than $0$, \url{b} should not be considered
as the path of a standard random walk.

Face to rare events, algorithmic and statistical approaches also
deliver similar conclusions. Let \url{c} denote a $100$-digit string
exclusively composed of $0$. To deduce whether \url{c} was generated
by tossing a coin, one can make 2 kinds of conjectures:
\begin{enumerate}
  \item according to computability theory, \url{c} has a relatively negligible Kolmogorov
complexity\footnote{As \url{c} can be easily generated with a short
command in most programming languages.} and can not be the outcome
of a coin tossing procedure.
\item in statistical terms, as the
probability of obtaining 100 successive ``heads'' (or ``tails'') by
tossing a coin is not more than $7.889e-31$, \url{c} is probably not
the result of such a procedure.
\end{enumerate}
However, despite the two preceding arguments, \url{c} could always
have been generated by tossing a coin, since even with a zero
probability, rare events do take place. This phenomenon is often
referred to as "black swans" by financial practitioners.

In this section, Kolmogorov complexity is presented by comparing it
with the statistical framework in regard to randomness
consideration. Two properties - invariance theorem and boundedness
theorem - of the former concept are stressed in this theoretical
introduction since they allow Kolmogorov complexity to be a good
indicator of the distance between a given sequence and random ones.

Based on this theoretical development, we will show in the following
sections how Kolmogorov complexity can be used in empirical data
analysis.

     \subsection{Kolmogorov complexity of price series: a basic example} \label{basique}
\begin{figure}[htbp]
        
         \begin{center}
           \includegraphics[scale = 0.35]{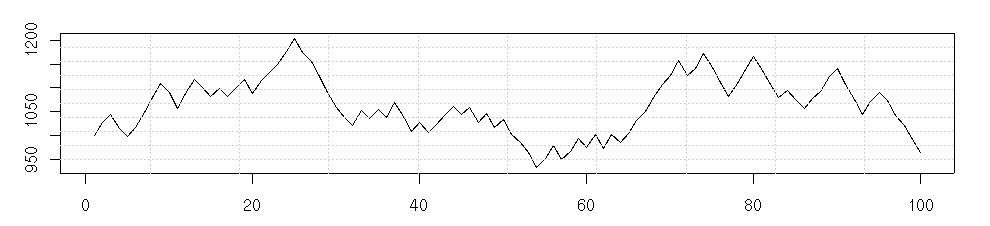}
            \caption{Simulated series $e_{1}$}\label{simple}
         \end{center}
         
       \end{figure}

After the theoretical introduction of Kolmogorov complexity, we
illustrate with the following example how to search structural
regularities in certain sequences behind their complex appearance.
This example is developed with a simulated price series which is
plotted in Figure \ref{simple}.

      The first 14 values of the simulated series are:
       \begin{center}

         $e_{1,t}=$\url{1000, 1028, 1044, 1015, 998, 1017, 1048, 1079, 1110, 1090, 1058, 1089,
           1117, 1100,...}
       \end{center}
     One apparent regularity in $e_{1,t}$ is that all prices seem to be concentrated around 1000. One can ``erase'' this regularity on taking the first order difference of $e_{1,t}$ with equation: $e_{2,t}=p_t-p_{t-1}$. This process delivers the following
sequence:
       \begin{center}
        $e_{2,t}=$ \url{28, 16, -29, -17, 19, 31, 31, 31, -20, -32, 31, 28, -17, -17...}
       \end{center}
    As positive series can be transformed into base 2 more easily, each
    element of $e_{2,t}$ is added by $32$, which gives:
       \begin{center}
           $e_{3,t}=$ \url{60, 48, 3, 15, 51, 63, 63, 63, 12, 0, 63, 60, 15, 15,...}
       \end{center}

      Then, $e_{3,t}$ is coded with binary numbers of 6 bits:
    \begin{center}
       $e_{4,t}=$\url{111100, 110000, 000011, 001111, 110011, 111111, 111111, 111111, 001100,
         000000, 111111, 111100, 001111, 001111,...}
    \end{center}
Without commas, $e_{4,t}$ becomes:
    \begin{center}
         $e_{5}=$\url{111100110000000011001111110011111111111111111111
         001100000000111111...}
    \end{center}
       Is  $e_{5}$ compressible? In other terms, remain-there structures in $e_{5}$ after the preceding transformations?
       Before answering these questions, one must notice that the binary expression of $e_{2}$ is already shorter than that of $e_{1}$. Our first transformation actually corresponds to a first compression by exploiting the sequential, concentrated aspect of $e_{1}$\footnote{$e_{1}$ requires 56 decimal digits for the first 14
prices, whereas $e_{2}$ only needs 30. Of course, we have to store
``1000'' somewhere in $e_{2}$ to keep a trace of the first price.
Measured in base 2, the length of $e_1$ turns from $4\times56=224$
bits into $4\times30=120$ bits, since each decimal digit ranging
from 0 to 9, is coded with at least 4 bits.}.

       Were $e_{5}$ incompressible, our ``regularity erasing procedure'' (thereafter, REP) would have attained its final stage. However, it is not the case. In $e_{5}$, there remains another structure which can be described as follows: if the $(2n-1)^{th}$ term is $1$ (resp. $0$), so will be the $(2n)^{th}$
term. By exploiting this regularity, $e_{5}$ can
       be compressed to $e_{6}$ which only carries every second element of $e_{5}$:
       \begin{center}\label{fin}
        $e_{6}$= \url{110100001011101111111111010000111110011011...}
       \end{center}

       At this step of REP, have we got an incompressible (random) string?
       Once again, this is not the case.  $e_{6}$ is actually produced by a pseudo-random generator which theoretically,
       can be reduced to its seed! Thus, $e_5$ is compressible in principle.
       However, it should be admitted that this ultimate compression is almost infeasible in practice, given the available range of compression tools.

       This basic example highlights how three important regularities \footnote{(i) Concentrated price values, (ii) particular structure in the binary digits and (iii) pseudo-random
       generator.} can hide behind the rather complex apparence of $e_{1,t}$, and how these structures can be revealed by a mere deterministic procedure.

       In the next sections, with the help of classical compression algorithms\footnote{We will use algorithms named Huffman, RLE and Paq8o8 which are presented in Annexe \ref{annexe4}.}, we show how REP can be implemented on both simulated (c.f. section \ref{artif}) and real-world financial time series (c.f. section \ref{reel}). In particular, we highlight that REP can be used to discover regularities actually undetectable by most statistical tests.
       Before concrete examples, we first advance some methodological considerations \ref{methodology}.

\section{A methodological point} \label{methodology}

The punch-line of this paper is to show how lossless compression
algorithms can be used to measure the randomness level of a numeric
string. This requires a series of special transformations. Initial
data must go through two consecutive operations: discretization and
REP.

To concentrate on the relation between compression and regularity,
in the last section, we show with a concrete example how REP can
reveal structures in integer strings. However, to apply this latter
procedure to real-world data, a discretization is necessary since
financial returns are often considered to be Real numbers, while
computer tools only work with discrete ones.

One can therefore query the effect of such discretization on
financial series' randomness. To answer this question, we introduce
the principle of lossless discretization.

\subsection{The lossless discretization principle}\label{uniforming}
In this section, we propose a generic methodology to estimate
Kolmogorov complexity of a logarithmic price series\footnote{The
reason why we use logarithmic price ($p_{t}$) here is that returns
($r_{t}$) in finance are usually defined as \emph{continuously
compounded returns}, \emph{i.e.} $r_{t}=ln(p_{t})-ln(p_{t-1})$}, as
what one can observe in real-world financial markets. As mentioned
in section \ref{basique}, a first ``compression'' consists of
transforming the logarithmic price sequence into returns. This
operation, familiar to financial researchers, actually delivers
unnecessarily precise data: for example, no one would be interested
by the $18^{th}$ decimal place of a financial return. Thus,
\textbf{without sensitive information loss}, one can transform real
number series into integer ones by associating to each integer a
certain range of real returns.

We also posit that integers used for discretization belong to a
subset of $\mathds{N}$ whose length is arbitrarily fixed to be
powers of 2. The total number of integers should be powers of 2,
since this allows to re-code the discretization result in base 2
without bias\footnote{Integer subsets fixed like this could be, for
example, from $0$ to $255$ or from $-128$ to $+127$.}. Once written
in binary base, financial returns become a sequence consisting of 0
and 1, which is perfectly suitable for compression tools.

After presenting the main idea of discretization, three points must
be discussed concerning this latter procedure:
\begin{enumerate}
\item Why not compress real returns in base 10 directly?\\
Had price variations been coded in base 10 and directly written in a
text file for compression, each decimal digit in the sequence would
have occupied 8 bits (one byte) systematically. As decimal digits
vary from 0 to 9 by definition, coding them with 8 bits would cause
a suboptimal occupation of stocking space: only 10 different bytes
would be used among the $2^8=256$ possibilities.\footnote{Each byte
can represent $2^8=256$ different combinations of 1 and 0. Thus, we
can potentially get $256$ different byes in a text file.} Thus, even
on random strings, spurious compressions would be observed because
of the suboptimal coding.\footnote{Low (used-bytes) /
(possible-bytes) rate.} Obviously, this kind of compression is of no
interest in financial series studies, and the binary coding system
is necessary to introduce Kolmogorov complexity in finance.

\item Is there any information loss when transforming real-number innovations into integers? Obviously, on attaching the same integer, for instance ``134'', to two close variations (such as ``1,25\%'' and ``1,26\%''), we reduce the precision of the initial data.\\
However, as mentioned above, the range of integers used in the
discretization process can be chosen from $[0, 255], [0,
511],\cdots,[0, 2^{n}-1]$ to obtain the ``right'' precision level:
neither illusorily precise nor grossly inaccurate.

This principle is shared by numerical photography: on choosing the
right granularity of a picture, we stock relevant information and
ignore some insignificant details or variations.

Following this principle, to preserve the fourth decimal place of
each return, one needs to use integers ranging from 0 to 8192.

\item If each return is coded with one single byte\footnote{This implies that Real-number returns are discretized with integers ranging from $0$ to $255$.}, discretized financial series can be transformed into \url{ASCII} characters.
 Lossless compression algorithms such as Huffman, RLE and PAQ8o8 can be used to search patterns in the corresponding text files.
\end{enumerate}


\subsection{REP: an incremental procedure for pattern detection}
As shown in section \ref{finance}, a generic methodology to capture
financial dynamics requires to identify the underlying random
process structuring $\varepsilon_t$ (for example, \emph{i.i.d.}
Gaussian random innovations).

On introducing Kolmogorov complexity in Finance, we want to consider
whether it is possible or not to find, besides the well-documented
stylized facts, other structures more subtly hidden in financial
returns. In other terms, we try to find new and fainter
regularities at the presence of more apparent ones.

If one uses compression methods on return series directly,
algorithms would only catch the ``main structures'' in the data, and
could potentially overlook weaker (but existing) patterns.

To expose these latters, we suggest an iterative process that, step
by step, erases the most evident structures from the data. Resulting
series are subject to compression tests for unknown structures. Due
to this process, even if financial returns witness obvious patterns
such as stylized facts, compression algorithms can always
concentrate on unknown structures, since once erased, identified
patterns would no longer be mixed with unknown structures, and any
further compression could be related to the presence of new
patterns.

To interpret the compression results from REP, we distinguish 3
situations:
\begin{enumerate}
\item The original series, denoted by $s$, is reduced to a perfectly determinist procedure that can be
actually detected by compression tools. In this case, $s$ is a
regular series.
\item $s$ is reduced to an incompressible
``heart''. In this case, $s$ is a random string.
\item $s$ has a determinist kernel, but this latter is too complexe to be exploited by compression tools in practice. In this case, $s$ should be considered as random.
For example, it is the case with normal series generated by a
programming language like ``R''\footnote{\citet{Rstats}}. REP's
first step is to erase the distribution law and to deliver a
sequence produced by the programming language's pseudo-random
generator. This generator, which is a perfectly determinist
algorithm, is a theoretically detectable and erasable structure.
Nevertheless, to our knowledge, no compression algorithm explores
this kind of structures. So, in practice, a binary series drawn from
an uniform law in ``R'' is considered as ``perfectly random'',
though this is not the case from a theoretical point of view.
  \end{enumerate}
Actually, given a finite time series, it is just impossible to
attest wether it is a regular or random one with certainty. At least
two arguments support this point:
\begin{enumerate}
\item As statistical conjectures are based on samples instead of populations, Kolmogorov complexity is estimated with finite strings.
Thus, none can tell \textbf{in a certain way} if a finite string is
a part of a random one.
\item  The function attaching to $s$ the true value of its Kolmogorov complexity cannot be calculated with a computer. In other terms, except in some very seldom case, one can never be sure if a given
program $P$ is the shortest expression of $s$.
\end{enumerate}
So, compression tools only \emph{estimate} Kolmogorov complexity,
none of them should be considered as an ultimate solution that will
detect all possible regularities in a finite sequence. This
conclusion is closely related to the G\"{o}del undecidability.

In this sense, compression algorithms share the same shortfall with
statistical tests: both approaches are inept to deliver certain
statements. Nevertheless, as statistical tests are making undeniable
contributions to financial studies, compression algorithms can also
be a powerful tool for pattern detection in return series.

To illustrate this point, we will compare algorithmic and
statistical methods in the next section with regard to structure
detection in numeric series.

\section{Non-equivalence between algorithmic and statistical methods}\label{artif}
In this section, we illustrate with stimulated data how compression
algorithms can be used to search regular structures. Statistical and
algorithmic tools are compared in these illustrations to show how
some patterns can be overlooked by standard statistical tests but
identified by compression tools.

To be more precise, with simulated data we will show:
\begin{enumerate}
  \item how statistical and algorithmic methods perform on random
  strings,
  \item how some statistically invisible structures can be detected by
  algorithmic tools,
  \item and that compression tests have practical limits also.
\end{enumerate}
        \subsection{Illustration 1: Incompressible randomness}\label{discret} 
       In this section, we simulate a series of 32000 $i.i.d.N(0,1)$ returns with the statistical language ``R'', and examine the simulated data with statistical tests and compression algorithms respectively.
        We show that compression algorithms cannot compress random strings, and sometimes they even increase the length of the initial data.

\begin{figure}[htbp]
                   \begin{center}
                     \includegraphics[scale = 0.25]{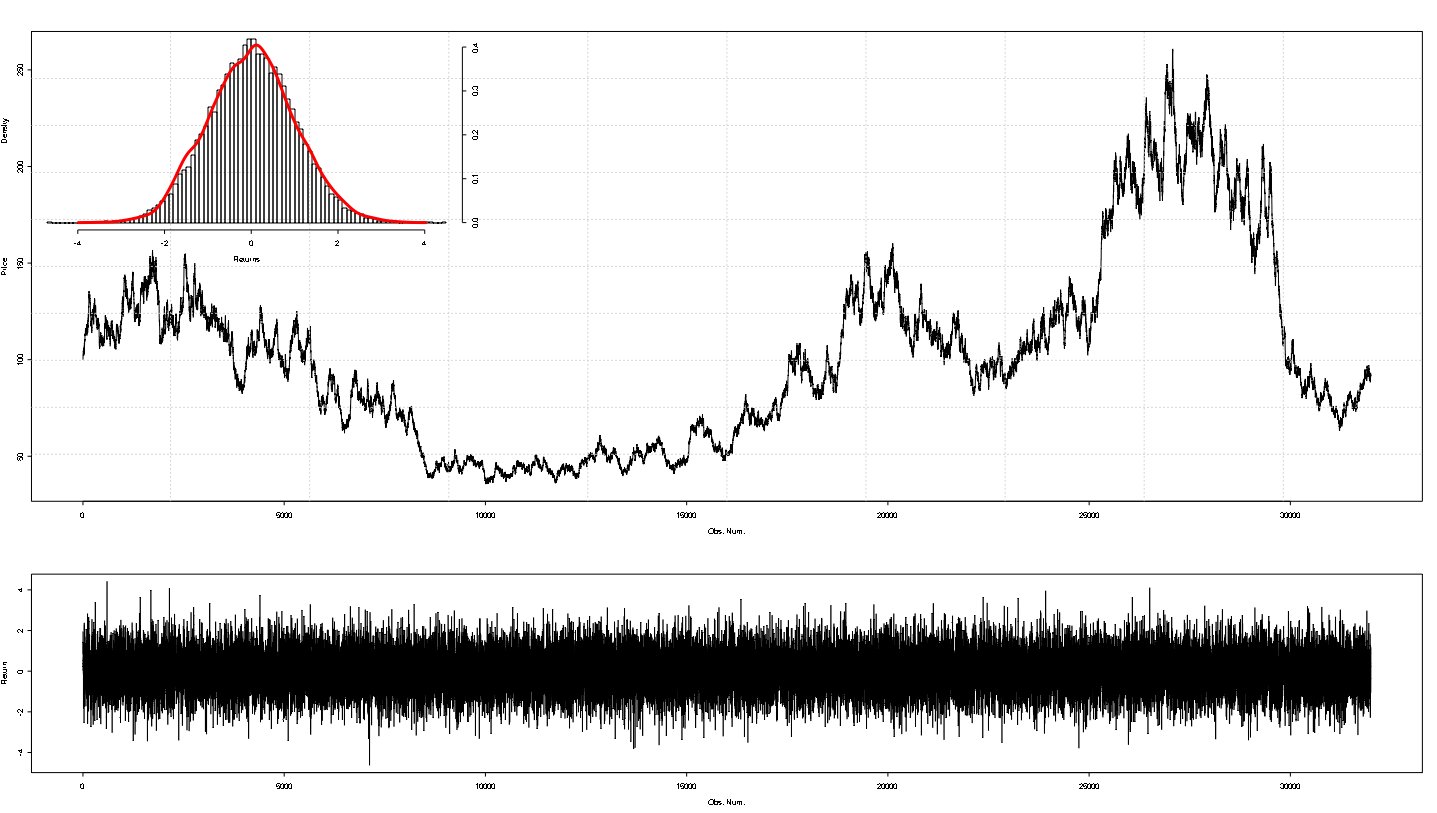}
                   \end{center}

                   \caption{Return series simulated from the standard normal distribution and the price motion corresponding to the simulated returns.} \label{baseline}

    \begin{scriptsize} Top-left: histogram of simulated returns.\\ Middle: price sequence generated from simulated returns, with an initial price of 100.\\
     Bottom: time series plot of simulated returns.\end{scriptsize}
                 \end{figure}

     The simulated series is quite similar to ``\emph{consecutive financial returns}'' and can be used to generate an artificial ``\emph{price sequence}''. Figure \ref{baseline} gives a general description of the data.

                 Statistical tests do not detect any specific structure in the normally distributed
                 series, as we can witness in Tables \ref{2}, \ref{1} and \ref{BDS_baseline}.

               \begin{table}[htbp]
                 \caption{Unit root test for simulated returns}\label{2}
                     \begin{center}
                      \begin{tabular}{p{3cm}| c |c| c}
                      Test&Val. statistical&order&\emph{p-value}\\
                      \hline
                      \hline
                      ADF&$-32.7479^{***}$&31&0.01\\
                      PP&$-178.52949^{***}$&16&0.01\\
                    \end{tabular}
                    \end{center}
                    \begin{center}
                     \small{$H_{0}$: the simulated series has a unit root.}
                  \end{center}
                \end{table}

                \begin{table}[htbp]
                  \caption{Autocorrelation tests for simulated returns }\label{1}
                       \begin{center}
                    \begin{tabular}{p{3cm}| c| c |c}
                      Series&$\chi-square$&deg. lib.&\emph{p-value}\\
                      \hline
                      \hline
                      baseline&0.1838&1&0.668\\
                      &36.9157& 36 &0.4264\\
                    \end{tabular}
                  \end{center}
                  \begin{center}
                  \small{$H_{0}$: the simulated series is $i.i.d.$.}\end{center}
                \end{table}

                 \begin{table}[htbp]
                  \caption{BDS test for simulated returns, \protect{$m=\{2,3\}$} }
                    \label{BDS_baseline}
                  \begin{center}
                    \begin{tabular}{c|c|c|c|c}
                      $\varepsilon$& 0.5012& 1.002& 1.5035& 2.0047\\
                      \hline
                      \hline
                      $m=2$&-0.2082&-0.2987&-0.5232&-0.7221\\
                      \emph{p-value}&0.8351&0.7651&0.6009&0.4702\\
                      \hline
                      $m=3$& 0.8351&0.7651&0.6009&0.4702\\
                      \emph{p-value} & 0.9503&0.9803&0.9344&0.8600\\
                    \end{tabular}
                  \end{center}
                  \begin{center}
                    \small{$H_{0}$: the simulated series is $i.i.d.$.}
\end{center}
                \end{table}

                 To check the performance of the algorithmic approach,
                 we implement compression tools on the simulated data. As explained in the last section, before using compression algorithms,
                 we should firstly transform the simulated real-number returns into integers (discretization).
                 In other terms, we must associate to
                 each return an integer ranging from 0 to 255, with ``0'' and ``255'' corresponding to the lower and upper bounds of the simulated data respectively.
                Here, instead of a "one to one" correspondence, each integer from 0 to 255
must represent a range of real numbers. The size of the interval
associated to each integer (denoted by $e$)
                 is fixed as follows:
                 \begin{equation}
                   e= (M-m)/256
                 \end{equation}
                 where $M$ and $m$ represent the upper and lower bounds of the simulated returns.

\begin{figure}[htbp]

                   \begin{center}
                     \subfigure[Regular bounds]{
                       \label{discretisation1}
                       \includegraphics[scale = 0.30]{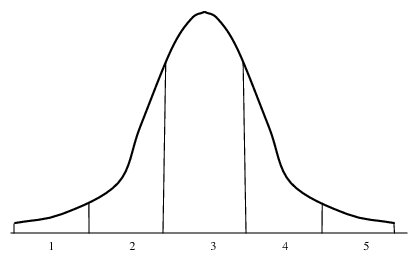}}
                     \hspace{.3in}
                     \subfigure[Identical probability]{
                      \label{discretisation2}
                       \includegraphics[scale = 0.30]{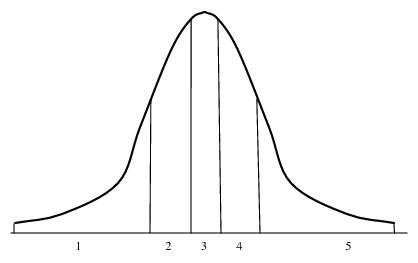}}
                     \caption{Data discretization: how to place bounds?}
                           \label{discretisation}
                   \end{center}
                 \end{figure}

                 Provided the value of $e$, we can divide the whole range $[m,M]$ into 256 intervals, and associate, to each return (denoted by $x$), an integer $k$ satisfying:
                 \begin{equation}\label{eqk}
                   x\in[m+(k-1) \times e, m+k \times e[
                 \end{equation}
                 In other terms, after sorting the 256 ``$e$-sized'' intervals in ascending order, $k$ is the rank of the one containing $x$. As $k$ varies from 0 to 255, it should be coded with 8 bits, since $2^8=256$.
                 Remark that, in dividing the real set $[m,M]$ into 256 identical subsets, we will obtain a series of regular bounds on X-axis under the standard normal distribution curve, as represented by Figure \ref{discretisation1}.
                 Using this discretization method,
                 we obtain normally distributed integers:
                 values close to
                 128 are much more frequent than those close to 0 or 255. An efficient
                 compression
                 algorithm will detect this regularity and compress the discretized data.

                 However, this normal-law-based compression is of little interest for financial
                 return analysis. According to REP, the
                 normal distribution should be erased from the
                 simulated data in order to expose more sequential structures.

                 To do this, we propose a second discretization process that will deliver uniformly distributed integers, instead of normally distributed ones.

                In this second process, the main idea stays the same: real number returns can be discretized in dividing the whole interval
                 $[m,M]$ into 256 subsets\footnote{Here, one can choose a power of 2 large enough to preserve all necessary information in the initial sequence. This is possible
since the initial data have a limited precision.}, and then in
associating to each return $x$ the rank of the subset containing
$x$.

                 However, this time, instead of getting equal-size subsets, we will fix the separating bounds in such a way that probability surface of the normal distribution would be divided into equal parts (c.f. Figure \ref{discretisation2}).
To be more precise, we want to define 257 real-number bounds,
denoted by $borne(0), borne(2), ..., borne(256)$, to make sure that
each value draw from $N(0,1)$ has the same probability (1/256) to
fall in each interval $[borne(i), borne(i+1)]$.
                 Subsets defined like this don't have the same
                 size: as illustrated in Figure
                 \ref{discretisation2},
                 the more a subset is close to zero, the smaller it
                 becomes.

               The discretization process described in Figure \ref{discretisation2} is implemented to simulated returns. Discretized returns are presented in Figure \ref{experience2_2}.
                The $i^{th}$ point in this figure represents the integer associated to the
$i^{th}$ return. One can notice that the plot of the discretized
returns is relatively homogeneous without any particularly dense or
sparse area (see Figure in appendix \ref{annexe5}).\footnote{It will
be shown that the plot of discretized returns is not always as
homogeny as this, especially when the initial data are not
\emph{i.i.d.}.} This rapid visual examination confirms the
                uniform distribution of discretized returns.

When lossless compression algorithms are used on the \url{ascii}
text obtained from the uniformly discretized returns (c.f. Figure
\ref{eperience2_3} in appendix \ref{annexe5}), these algorithmic
tools turn out to be inefficient (c.f. Table \ref{compr_exp3}).

                \begin{table}[htbp]

                  \begin{center}
                    \caption{Compression tests}
                    \label{compr_exp3}
                    \begin{tabular}{p{3cm}| c c}
                      Algorithm&file size&compression rate\\
                      \hline
                      \hline
                      &32000&0\%\\
                      Huffman&32502&-1.57\%\\
                      Gzip & 32073 & -0.23\%\\
                      PAQ8o8&32118&-0.37\%\\
                    \end{tabular}
                  \end{center}
                  \begin{center}
                   \small{Interpretation: \emph{discretized returns seem to be
incompressible}}
                  \end{center}
                \end{table}

                This example clearly shows that normally distributed returns can be transformed into an
                uniformly distributed integer string whose length is entirely irreducible by lossless compression tools. We can conjecture from this
                experiment, that the initial series has a Kolmogorov complexity close to its length.

                With data simulated from the standard normal law, we
                show that statistical and
                algorithmic methods deliver the same conclusion on random strings.

               In a further comparison between the 2 approaches, we will "hide" some structures in an uniformly distributed sequence,
                and show that the hidden regularity is detectable by compression algorithms but not by statistical tests.

                \subsection{Illustration 2: Statistically undetectable structures and compression algorithms}
                In this section, we make a further comparison between statistical and algorithmic methods with simulated series.
                More precisely, instead of random strings, we generate structured data to check the power of these two methods in pattern detection.

                A return series can carry many types of structures.
                Some of them are easily detectable by standard
                statistical tests (for example, auto-regressive process or
                conditional variance process). However, to distinguish compression algorithms from statistical tests, this kind of
                regularities will not be the best choice. Here, we want to build statistically undetectable regularities that can be revealed by compression tools.

               In this purpose, the return series is simulated as follows:
                \begin{enumerate}
                \item Draw $32000$ integers from the uniform law $U(0,255)$\footnote{$U(0,255)$ denotes the }, and denote by $text$ the sequence containing these integers. $text$ is then submitted to several
                transformations to ``hide'' statistically undetectable regularities behind its random appearance.

                Let $text'$ denote the biased series which should be distinguished
                from the uniformly distributed one, $text$.

                $text'$ is obtained from $text$ by changing the
                last digit (\emph{resp.} last 3 digits) of the binary
                expression of each integer in $text$.

                To be more precise, in case 1, $text'$ exhibits an alternation between $0$ and $1$ on the last digit of each term.
                In case 2, elements from $text'$ repeat the cycle $000, 001, 010, 011, 100, 101, 110, 111$ on the last 3 digits.

               For example, in coding each integer in $text'$ with 8 bits, we could get, in case 1,
               a sequence as follows:
                $$\{0000000\underline{1}, 00011011\underline{0},
                1110100\underline{1},
                10000111\underline{0}, 1000011\underline{1}, ...\}$$
                This regularity is actually a "parity alternation", since binary numbers ended by $1$ (\emph{resp.} $0$) are always odd (\emph{resp.} even).

                In case 2, we could get a sequence like: $$\{01010\underline{000},
                11101\underline{001}, 01101\underline{010},
                01101\underline{011}, ...,
                11100\underline{111}, 00011\underline{000}\}$$

              \item The biased integer sequence $text'$ is what we want to obtain after the discretization of real number returns. So, the second
              step of our simulation process is to transform
              $text'$ into return series.

              In other terms, we should associate to each integer in $text'$ a real number return.
              This association is based on the separating bounds calculated in the last section (c.f. $borne(0)$, $borne(2)$, $\cdots$, $borne(256)$ described in paragraph \ref{discret}).
              To each integer in $text'$, denoted by $text'[i]$, we attach a real number that is independently drawn from the uniform law $U(borne(text'[i]), borne(text'[i]+1))$,
              Let $chron$ denote the return series obtained from this transformation.

               By construction, $chron$ has two properties:
                    \begin{enumerate}
                    \item globally, its terms follow a normal law;
                    \item if we discretize $chron$ with the process described in paragraph \ref{discret}, the resulted integer sequence will be exactly $text'$.
                    \end{enumerate}
\end{enumerate}
                So, by construction, $chron$ is a normally
                distributed series that will exhibit patterns after discretization. Are these structures detectable by both
                statistical and algorithmic methods? Or only one of these approaches can reveal the hidden regularities?
                 That's what we want to see with the
                 following tests.

\begin{figure}[htbp]
                                   \begin{center}
                    \subfigure[Case 1]{
                      \label{exper_3}
                      \includegraphics[scale = 0.125]{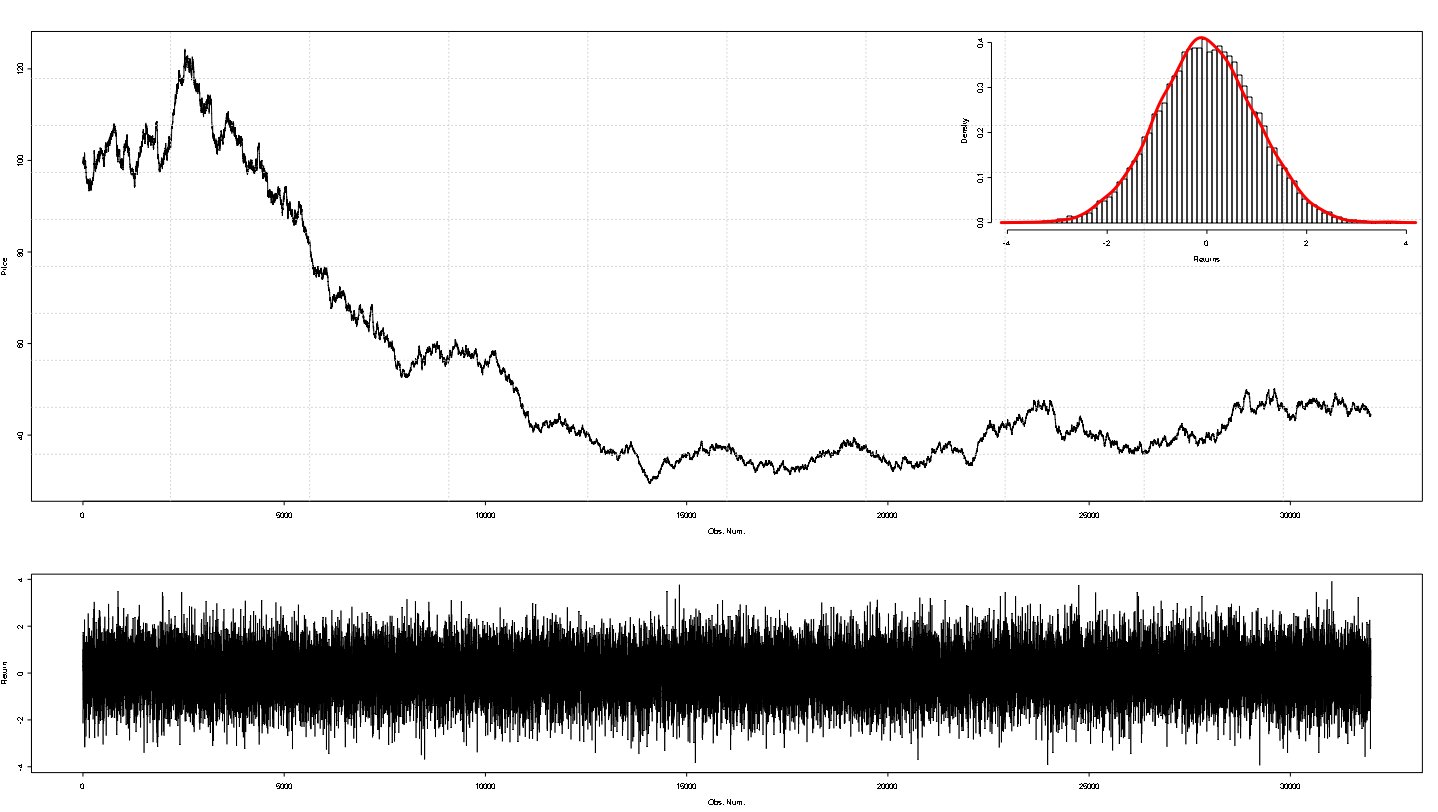} }
                    \hspace{.3in}
                    \subfigure[Case 2]{
                       \label{exper_4}
                      \includegraphics[scale = 0.125]{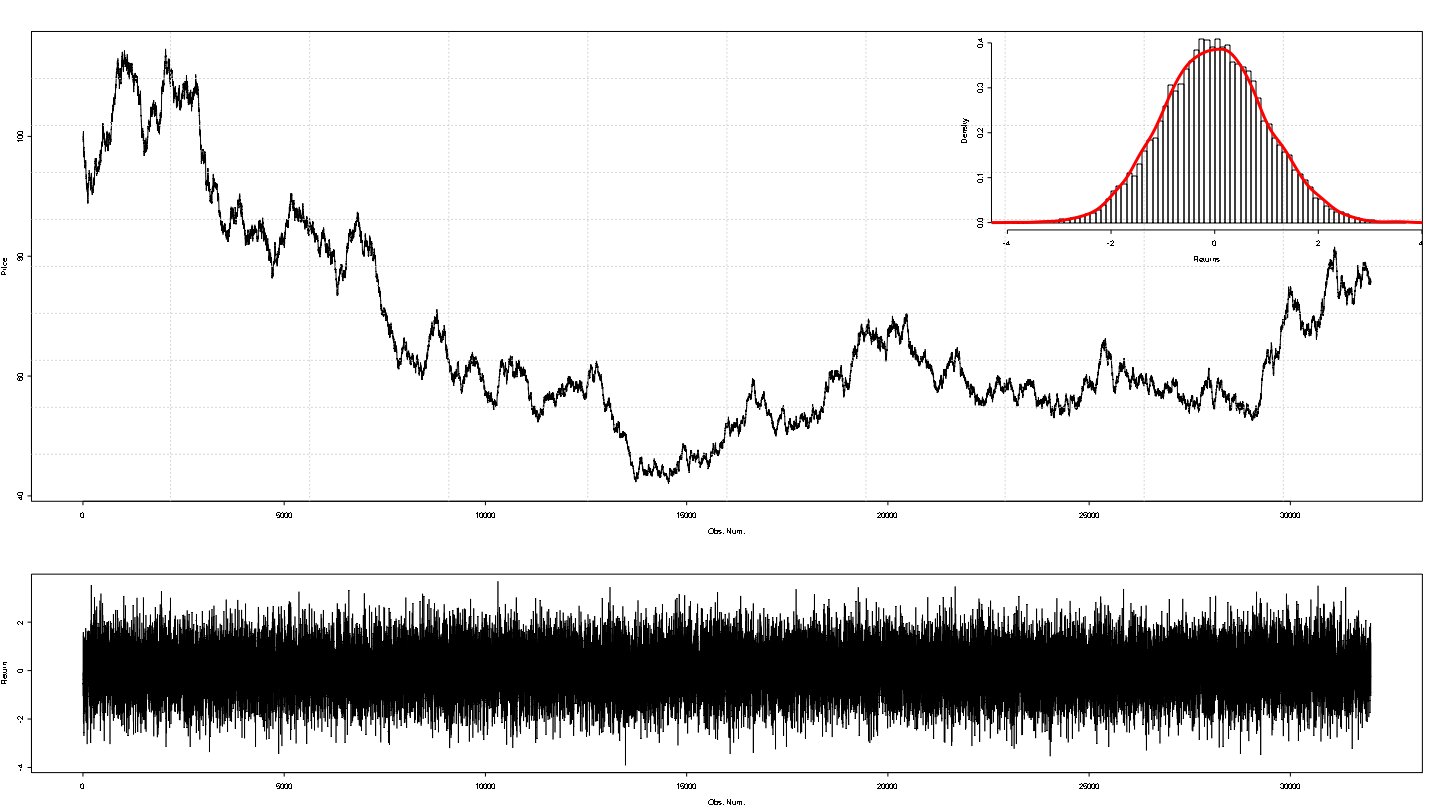}}
                    \caption{Simulated returns with two important structures. On the top of each figure, we plotted the pseudo price series obtained from $chron$ with an initial price of 100.
                    At the bottom, we plotted the simulated return series $chron$.} \label{exper_34}
                  \end{center}
                \end{figure}

                Simulated returns are plotted in Figure
                \ref{exper_34}, with Figure \ref{exper_3} corresponding to case 1, and Figure \ref{exper_4} to case 2.

           As shown in Tables \ref{uroot_pairimpair}, \ref{autocor_pair_impair} and \ref{BDS_pairimpair}, statistical tests detect no
           structure in $chron$.
     \begin{table}[htbp]
                      \begin{center}
                        \caption{Unit root tests for $chron$}   \label{uroot_pairimpair}
                        \begin{tabular}{p{3cm}| c c c| c c c}
                          &\multicolumn{3}{c}{case 1}&\multicolumn{3}{c}{case 2}\\
                          Test&Val. statistic&order&\emph{p-value}\\
                          \hline
                          \hline
                          ADF&$-31.5598^{***}$&31& 0.01&$-31.2825^{***}$&31&0.01\\
                          PP& $-178.797^{***}$&16&0.01&$ -179.9449^{***}$& 16& 0.01\\
                        \end{tabular}
                      \end{center}

                          \begin{center}

                     \small{$H_{0}$: $chron$ has a unit root.}
                  \end{center}
                    \end{table}

                \begin{table}[htbp]
                  \begin{center}
                    \caption{Autocorrelation tests for $chron$} \label{autocor_pair_impair}
                    \begin{tabular}{c c c| c c c}

                      \multicolumn{3}{c}{case1}&\multicolumn{3}{c}{case2}\\
                      \hline
                      $\chi-square$&deg. lib.&\emph{p-value}&$\chi-square$-&deg. lib.&\emph{p-value}\\
                      \hline
                      \hline
                      0.0096&1&0.9219&1.1169& 1&0.2906\\
                      29.6655& 36 &0.7629&45.4802&36&0.1337\\
                    \end{tabular}
                  \end{center}
                  \begin{center}
  \small{$H_{0}$: $chron$ is not autocorrelated.}
                  \end{center}

                \end{table}

                \begin{table}[htbp]
                  \caption{BDS tests for $chron$, \protect{$m=\{2,3\}$} }\label{BDS_pairimpair}
                  \begin{center}
                    \begin{tabular}{c|c|c|c|c||c|c|c|c|c}
                      \multicolumn{5}{c}{case 1}&\multicolumn{5}{c}{case 2}\\
                      \hline
                      $\varepsilon$&0.5006& 1.0012& 1.5019& 2.0025&$\varepsilon$& 0.4988& 0.9976& 1.4964& 1.9952 \\
                      \hline
                      \hline
                      $m=2$& 0.1121&0.1790 &    0.3377&     0.4318&$m=2$&0.1886  &   0.1186  &   0.0627  &   0.1207 \\
                      \emph{p-value}& 0.9108  &   0.8579&     0.7356
                      &  0.6659&\emph{p-value}& 0.8504    & 0.9056
                      &  0.9500 &    0.9039 \\
                      \hline
                      $m=3$& 0.1329 &    0.2233  &   0.3578 &
                      0.4662&$m=3$& -0.0974  &  -0.0617   &  0.0336 &
                      0.2153 \\
                      \emph{p-value} &0.8943  &   0.8233 &    0.7205
                      &    0.6411&\emph{p-value} & 0.9224  &   0.9508
                      &   0.9732  &   0.8295\\
                                  \hline
                    \end{tabular}
                  \end{center}
                  \begin{center}
                     \small{$H_{0}$: $chron$ is $i.i.d$.}
                  \end{center}
                    \end{table}

                    After statistical tests, the REP is applied to $chron$. From a theoretical point of view, regularity hidden in case 1 implies a
                    compression rate of $12.5\%$.
                    This rate is calculated as follows: each integer in the discretized $chron$ (which is actually $text'$) is coded with 8 bits, while only 7 of them are necessary. Actually, given the "parity alternation", the last digit of each byte in $text'$ is determinist.
                    In other terms, we can save 1 bit on every byte. Whence the theoretical compression rate
                    $1/8=12.5\%$.
                    Following the same principle, we can calculate the theoretical compression rate in case
                    2: $3/8=37.5\%$.

                    Table \ref{compress_pairimpair1} presents
                    compression results in the 2 cases. Notice that
                    realized compression rates are close to theoretical ones, but they never attain their exact value. Among
                    the three algorithms in use, $Paq8o8$
                    offers the best
                    estimator of theoretical rates.

                    \begin{table}[htbp]
                      \begin{center}
                        \caption{Compression tests}\label{compress_pairimpair1}
                        \begin{tabular}{p{3cm}| c c | c c}
                          &\multicolumn{2}{c}{case 1}&\multicolumn{2}{c}{case 2}\\
                          Algorithm&file size&compression rate&file size&compression size\\
                          \hline
                          \hline
                              & 32000 & 0\%& 32000  & 0\%\\

                          Huffman&31235&2.39\%&23079&27.88\%\\
                          Gzip & 31322 & 2.12\%& 23160  & 27.63\%\\
                          PAQ8o8&28296&11.58\%&20974&34.46\%\\
                        \end{tabular}
                      \end{center}

                      \small{Interpretation in case 1 : \emph{$text'$ is compressible}.}\\
                      \small{Interpretation in case 2 : \emph{$text'$ is compressible}.}
                    \end{table}

                    In this illustration, we show that the algorithmic approach,
                    essentially based on Kolmogorov complexity,
                    can sometimes identify statistically-undetectable structures in simulated data.

                    However, as mentioned in the theoretical part, the ultimate algorithm calculating the true Kolmogorov complexity for all binary
                    strings \textbf{does not} exist. Compression tools also have
                    practical limits. In other terms, certain structures are undetectable by available compression tools.
                    We show this point in the next section.
        \subsection{Practical limits of the algorithmic method: decimal digits of $\pi$}
\begin{figure}[htbp]
                   \begin{center}
                     \includegraphics[scale = 0.25]{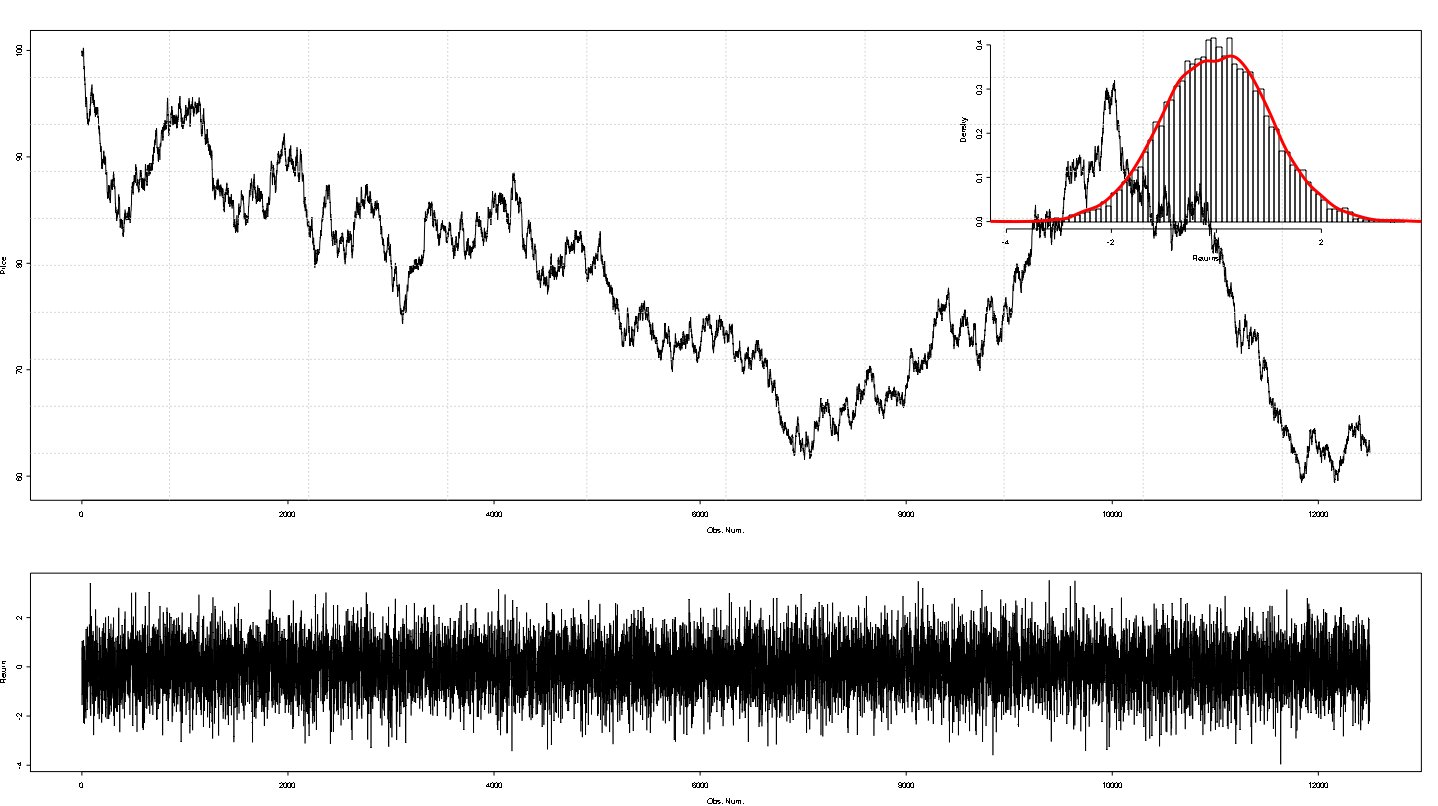}
                   \end{center}
                   \caption{Pseudo financial time series generated from decimals of $\pi$}\label{pi1}
                 \end{figure}

                Besides the Euler numbers and the Fibonacci numbers, $\pi$ is perhaps one of the most studied mathematic
                numbers. There are many methods to calculate $\pi$.
                The following two equations both deliver a big number of its
                decimal digits:
\begin{itemize}
                \item Leibnitz-Madavar's formula,
                \begin{equation}
                  4 \times \sum_{n=0}^{\infty}\frac{(-1)^n}{2n+1}
                \end{equation}
                \item And the second formula :
                \begin{equation}
                  \pi=\sqrt{6\times(1+\frac{1}{2^2}+\frac{1}{3^2}+\frac{1}{4^2}+...+\frac{1}{n^2})}
                \end{equation}
                 \end{itemize}
               $\pi$ can be transformed into a return series in 3 steps:
                \begin{enumerate}
                  \item Each decimal digit of $\pi$ is coded in base 2 with 4 bits. For example, the first 4 decimals (c.f. $1, 4, 1, 5$) become $0001$, $0100$, $0001$, $0101$. Following this principle, the first 50000 decimals of $\pi$ correspond to 200000 bits of binary information.
                  \item The 200000-digit binary string obtained from the first step is then re-organized in bytes. For example, the first 4 decimals make two successive bytes: $00010100$, $00010101$. Each byte corresponds to a integer ranging from 0 and 255. Here, the first two bytes of $\pi$ become 20 and
                  21. Denote by $\pi'$ the integer sequence after this re-organization.
                  \item Finally, we associate a real-number return to each integer in $\pi'$. To do this, we follow the same principle as
                  in the last section: to each term of $\pi'$, denoted by $\pi'_t$
                  ($t\in[1,25000]$), we associate a real number
                  that is independently drawn from the uniform distribution
                  $U(born(\pi'_t), born(\pi'_t+1))$. Where $borne(i)$ ($i\in[0,256]$) means the $i^{th}$ separating bound obtained from the uniform discretization of the
                  normally distributed return series in section \ref{discret}. After this step, we get a pseudo return series plotted in Figure
                  \ref{pi1}.
                \end{enumerate}

                 As exposed in Table \ref{compress_pairimpair}, $\pi$-based return series is incompressible
after discretization, since to our knowledge, no compression
algorithm exploits decimals of $\pi$.

                 The $\pi$-based simulation is another example that shows the possibility to hide patterns behind a random appearance.
                 It also witnesses that some theoretically compressible structures may be overlooked by available compression tools.
                  These structures are perfectly compressible in theory, but not in practice yet.

                \begin{table}[htbp]
                \begin{center}
                  \caption{Compression test: $\pi$} \label{compress_pairimpair}
                  \begin{tabular}{p{3cm}| c c }
                    &\multicolumn{2}{c}{case 1}\\
                    Algorithm&file size&compression rate\\
                    \hline
                    \hline
                    & 12500 & 100\%\\

                    Huffman& 12955& -3.64\%\\
                    Gzip & 12566 & - 0.528\%\\
                    PAQ8o8& 12587& -0.70\%\\
                  \end{tabular}
                \end{center}
                \begin{center}
               \small{Interpretation: \emph{We can't compress the $\pi$ based return series.}} \end{center}
              \end{table}
To check the performance of statistical tools on decimals of $\pi$,
we conducted the same tests as in the preceding illustrations. We
notice in tables \ref{uroot_pi} \ref{autocor_pair_pi} and
\ref{BDS_pi}, that statistical tests do nothing better than
compression tools: none of them can reject $H_0$ which implies the
absence of regularity.
 \begin{table}[htbp]
                \begin{center}
                  \caption{Unit root test for the series constructed by $\pi$} \label{uroot_pi}
                  \begin{tabular}{p{3cm}| c c c}
                    Test&Val. statistical&order&\emph{p-value}\\
                    \hline
                    \hline
                    ADF&$-23.3799^{***}$&23& 0.01\\
                    PP& $-110.1364^{***}$&13&0.01\\
                  \end{tabular}
                \end{center}

                \begin{center}

                     \small{$H_{0}$: the $\pi$ based series has a unit root.}
                  \end{center}
               \end{table}

              \begin{table}[htbp]
                \begin{center}
                  \caption{Autocorrelation tests}\label{autocor_pair_pi}
                  \begin{tabular}{c c c}
                    $\chi-square$&deg. lib.&\emph{p-value}\\
                    \hline
                    \hline
                    2.7339&1& 0.09824\\

                  \end{tabular}
                \end{center}
                \begin{center}

                     \small{$H_{0}$: the $\pi$ based series is not autocorrelated.}
                  \end{center}

              \end{table}

              \begin{table}[htbp]
                \caption{ BDS test for the $\pi$ based series, \protect{$m=\{2,3\}$} }\label{BDS_pi}
                \begin{center}
                  \begin{tabular}{c|c|c|c|c}
                    $\varepsilon$&0.5023 &1.0046 & 1.5069&2.0092 \\
                    \hline
                    \hline
                    $m=2$& -0.0895&0.0468&0.0395&0.0129\\
                    \emph{p-value}&0.9287&0.9627&0.9685&0.9897\\
                    \hline
                    $m=3$& -0.4005 &   -0.1755 &   -0.2780 &   -0.3208\\
                    \emph{p-value} &  0.6888   &  0.8607  &   0.7810 &    0.7483\\
                  \end{tabular}
                \end{center}
                     \begin{center}
                     \small{$H_{0}$: the $\pi$ based series is $i.i.d$.}
                  \end{center}
              \end{table}

In this section, simulated data are used to illustrate the
performance of compression tools in pattern detection.
 Two main results are supported by these illustrations: (1) Some statistically undetectable patterns can be traced by compression
 tools. (2) Certain structures stay undetectable by currently available compression
 tools.

 In the next section, we test the algorithmic approach with real-world financial return series.

\section{Kolmogorov complexity of real-world financial data: the case of Dow Jones industrial index} \label{reel}

\begin{figure}[htbp]
                   \begin{center}
                     \includegraphics[scale = 0.25]{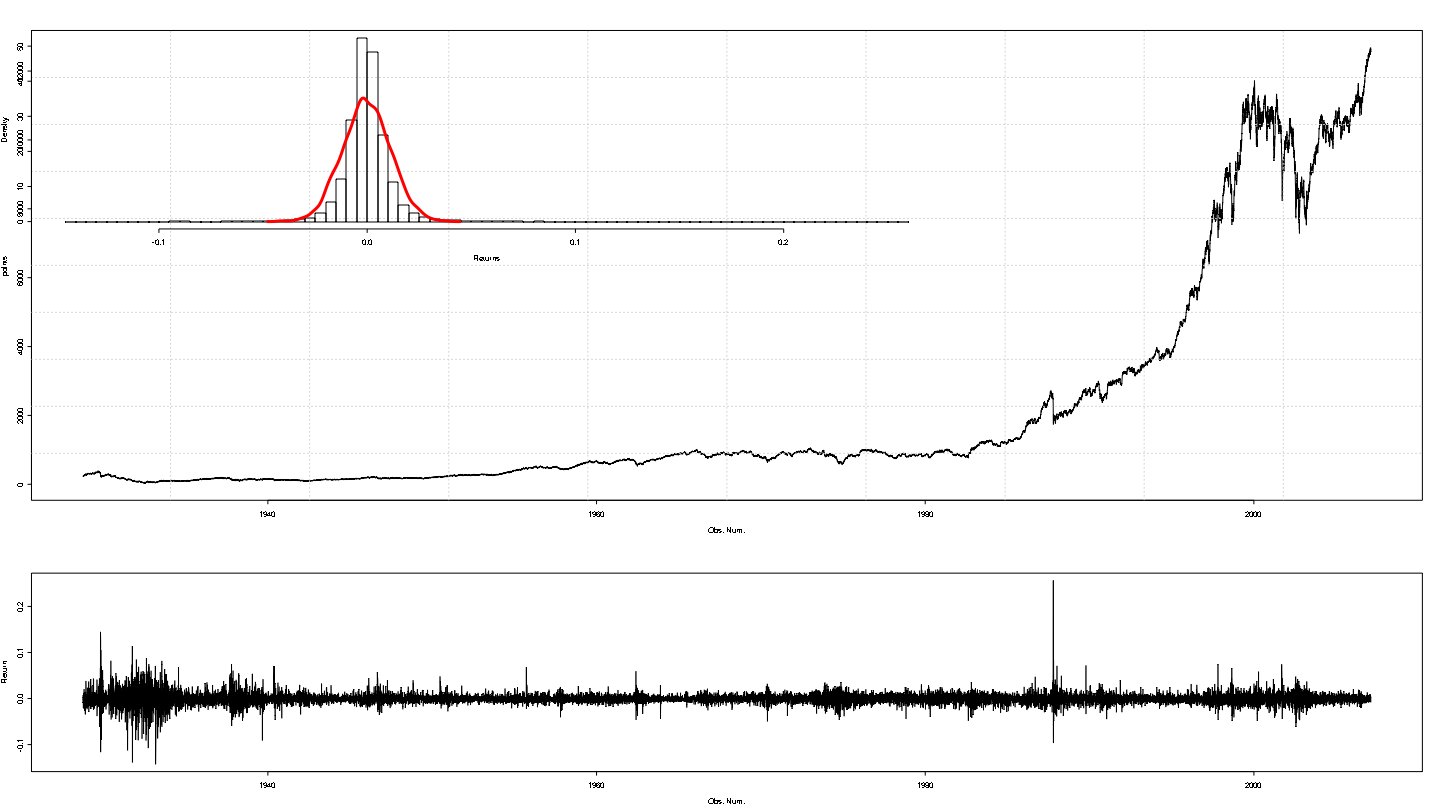}
                   \end{center}
                   \caption{Series constructed from Dow Jones daily closing prices.}
                                \label{dowdayly}
                                \scriptsize \hspace{4.4cm} Top-left: histogram of logarithmic differences.

                                \scriptsize \hspace{4.4cm} Middle: Dow Jones daily closing prices observed from 01/02/1896 to 30/08/2005.

     \scriptsize \hspace{4.4cm} Bottom: time series plot of the real-number returns.
                            \end{figure}

      In this section, we estimate the Kolmogorov complexity of real-world financial returns with lossless compression algorithms.
      To do this, we use the logarithmic difference of Dow Jones daily closing prices observed from
      01/02/1896 to 30/08/2005. Data used in this study are extracted from Datastream.
      Our sample containing 27423 observations is plotted in Figure \ref{dowdayly}.

Following the REP, we uniformly discretize the real-number returns
to prepare them for compression tests.

While separating bounds that are used to discretize simulated data
in the preceding sections all come from the standard normal law,
real-world returns cannot be tackled in the same way, since it is
well known that financial returns are not normally distributed.
Actually, there is no consensus on the way financial returns are
distributed in the financial literature. Therefore, to discretize
Dow Jones daily returns, separating bounds ($borne(i)$) should be
estimated from their empirical distribution.

                  Such an estimation can be realized in 3 steps:
                  \begin{enumerate}
                   \item sort the whole return series in ascending
                   order,
                   \item divide the ascending sequence into 256
                   equally-sized
                   subsets,
                  \item each return is represented by the rank of the subset containing it.\end{enumerate}
                 The advantage to estimate $borne(i)$ with this 3-step approach is that one can discretize a sample without
                 making any hypothesis on the distribution law of the population.

\begin{figure}[htbp]
                   \begin{center}
                     \includegraphics[scale = 0.25]{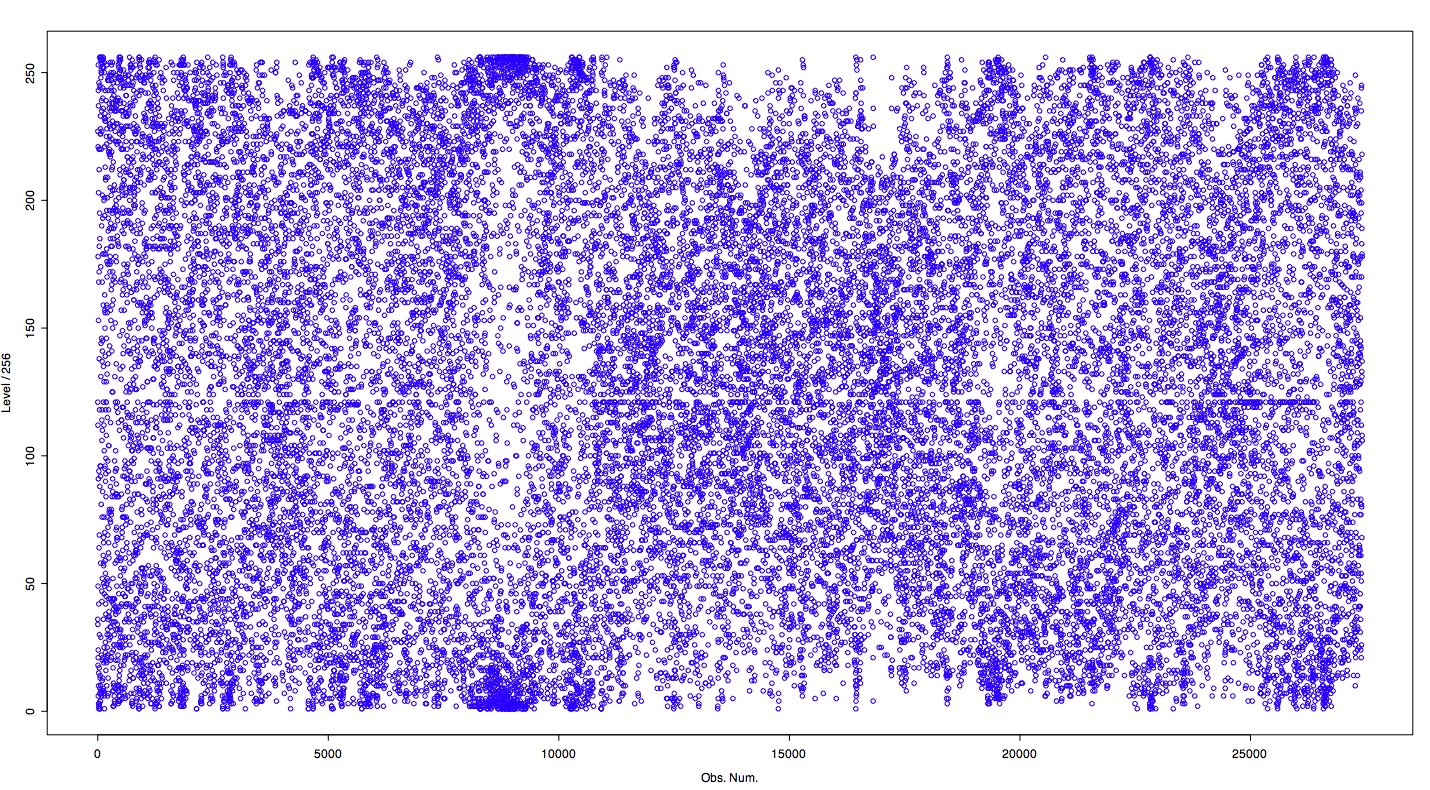}
                   \end{center}
                   \caption{Uniformly discretized Dow jones daily returns} \label{unidj}
                 \end{figure}

Figure \ref{unidj} is a plot of discretized Dow Jones daily returns.

One can notice in this figure that the uniform discretization fails
to deliver a perfectly homogeneous image as in the case of normally
distributed returns. Several areas seem to be sparser than the
others. For instance, regions exposing returns from Time 9000 to
Time 10000 and from Time 13000 to Time 20000 both present a great
difference in point density.

This statement can be explained by the volatility clustering
phenomenon, a well documented stylized fact in Finance. For example,
from Time 9000 to Time 10000, extreme values are obviously more
frequent than those close to 0.

This eye-detectable structure is confirmed by compression tests. As
we can witness in Table \ref{compress_dow uniformise}, algorithm
PAQ8o8 obtains a $0.82\%$ compression rate on the uniformly
discretized series.


\begin{table}[htbp]

     \begin{center}
       \caption{Compression tests on discretized DJ}
       \label{compress_dow uniformise}
               \begin{tabular}{p{3cm}| c c }
         Algorithm&file size&compression rate\\
         \hline
         \hline
         & 27423 & 100\%\\

         Huffman&27456 &-0.12\%\\
           Gzip &27489&-0.24\% \\
         PAQ8o8&27198 &0.82\%\\
       \end{tabular}
     \end{center}
\begin{center}
     \small{Interpretation: \emph{Discretized Dow Jones is compressible by PAQ8o8}}
\end{center}
   \end{table}



This compression rate appearing extremely weak, its robustness may
appear doubtful. To test its significance, we simulate 100 integer
sequences from an $i.i.d.U(0, 255)$ process, with each of them
containing 27423 observations as the Dow Jones daily return series
(thereafter DJ). These simulated sequences are tested by compression
tools. PAQ8o8 delivers no positive compression rate on none of the
simulated series.

This result confirms the theoretical relation between regularity and
compressibility: despite its weak level, the compression rate got
from discretized DJ indicates the presence of volatility clusters in
the data.

To advance another step in the REP and verify wether or not daily DJ
returns contain other patterns than the witnessed stylized facts, we
should remove the ``volatility clustering'' phenomenon with the help
of reversible transformations.

In fact, in the uniform discretization process described above, each
integer represents the same range of real returns. That is why
highly (\emph{resp.} lowly) volatile periods are marked by an
over-presence (\emph{resp.} a sub-presence) of extreme values in
Figure \ref{unidj}. To remove ``volatility clusters'' is to modify
these heteronomous areas and to ensure that each part of Figure
\ref{unidj} has the same point density.

The solution we propose here is to discretize DJ in a progressive
way. To be more precise, instead of discretizing the entire series
at once, we treat it block by block with an iterative procedure.

\begin{figure}[htbp]
                   \begin{center}
                     \includegraphics[height=8cm, width=16cm]{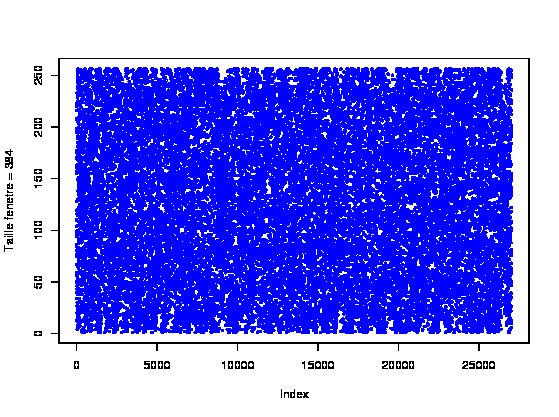}
                   \end{center}
                   \caption{DJ after the progressive discretization}  \label{DJsansBV}
                 \end{figure}

Denote by $S_t$ the Dow Jones daily return series, the following
3-step procedure can erase clustering volatilities in $S_t$:
\begin{itemize}
\item To start up, a 512-return sliding window is placed at the beginning of $S_t$. Returns in
the window (i.e. the first 512 ones) are transformed into integers
with the above-described unform discretization procedure. The
integer associated to $S_{512}$ is be stocked as the first term of
the discretized sequence.

\item The sliding window moves one step to the right, returns in the window are discretized again, and the integer
corresponding to $S_{513}$ is stocked as the second term of the
discretized sequence.

\item Repeat the second step until the last return of $S_t$.
The integer sequence obtained from this procedure doesn't reveal any
volatility cluster. A comparison between Figures \ref{DJsansBV} and
\ref{unidj} illustrates this progressive procedure's impact.
\end{itemize}

How this iterative process erases volatility clusters? The main idea
is to code each return in DJ with those appearing in its close
proximity. In the above example, the sliding window's length is
fixed to 512, this implies that each return's discretization only
depends on the 511 preceding terms.

Therefore, each integer in the discretized sequence, for example
$255$, can represent a $10\%$ rise during highly volatile periods as
well as a $3\%$ up during less volatile ones. Due to the progressive
discretization, extreme values in volatile periods are ``pulled''
back to zero, and returns during less volatile periods will be
``pushed'' to extreme values.

Were the non-uniform distribution and volatility clusters the only
regularities in DJ, the progressively discretized sequence (i.e.
$s_t$) would be incompressible by algorithmic tools. Positive
compression rates obtained on $s_t$ would indicate the presence of
unknown structures.

To check this, compression tests are conducted on $s_t$. Results are
presented in Table \ref{DJ F384}:
\begin{table}[htbp]

     \begin{center}
       \caption{Compression test: DJ after progressive discretization}
       \label{DJ F384}
               \begin{tabular}{p{3cm}| c c }
         Algorithm&file size&compression rate\\
         \hline
         \hline
         & 27039 & 100\%\\

         Huffman&27075&-0.12\%\\
         Gzip &27105&-0.24\% \\
         PAQ8o8&26913 &0.27\%\\
       \end{tabular}
     \end{center}
     \begin{center}
     \small{Interpretation: \emph{Even after the progressive discretization process, DJ remains compressible by PAQ8o8.}}
\end{center}
   \end{table}
We notice in this table that even after the progressive
discretization process, DJ remains compressible by PAQ8o8. This
result could indicate the presence of unknown structures in
financial returns. To better understand these structures, further
research is necessary to identify their nature and tell how to
remove them from $s_t$ and advance once again in the REP.

However, as we can witness in Table \ref{DJ F384}, compression rate
based on these unknown structures is extremely weak (c.f. $0.27\%$).
This indicates a high similitude between $s_t$ and a random string.
In other terms, although not completely random, once stylized facts
erased, Dow Jones daily returns have an extremely high Kolmogorov
complexity.

 In a certain degree, this result supports the EMH like most statistical works in
 Finance (see for eg. \citet{Lo2006}). Once unprofitable stylized facts are erased from
 DJ, the latter series is quite similar to a random string. The high Kolmogorov complexity observed in our study indicates to which extent it is difficult to
 find a practical trading rule that is outperforming the market in the long run.

\section*{Conclusion}

In this paper, we propose a generic methodology to estimate the
Kolmogorov complexity of financial returns. With this approach, the
weak-form efficiency assumption proposed by \citet{Fama1970} can be
studied by compression tools.

We give examples with simulated data that illustrate the advantages
of our algorithmic method : among others, some regularities that
cannot be detected with statistical methods can be revealed by
compression tools.

Applying compression algorithms to daily returns of the Dow Jones
Industrial Average, we conclude on an extremely high Kolmogorov
complexity and by doing so, we propose another empirical observation
supporting the impossibility to outperform the market.

A limit of our methodology lies in the fact that currently available
lossless compression tools, initially developed for text files, are obviously
not particularly designed for financial data. They could consequently
be restricted in their use for detecting patterns in financial motions. Future
researches could therefore develop finance-oriented compression tools to
establish a more direct link between a compression rate delivered from
a return series and the possibility to use any possible hidden pattern to outperform a simple buy and hold strategy.

Another important point to be stressed in this paper lies in the
fact that our methodology proposes an iterative process which will
improve, step by step, our comprehension on the "stratified" structure of financial price motions (\emph{i.e.} layers of mixed structures).

As illustrated in the empirical part of this research, even after
removing some of the more evident stylized facts from real-world
data, compression rates obtained from the Dow Jones daily returns
seem to indicate the presence of unknown structures. Although these
patterns could be difficult to exploit by strategy designers, at
least from a theoretical point of view, it is challenging to
understand the nature of these unknown structures. The next step
could then consist in removing them from the initial data, if
possible, and go one step further in the iterative process seeking
at identifying new regularities. The ultimate goal of this process,
even if it is probably a vast and perhaps quixotic project, could be
to obtain an incompressible series, and, so to speak, to reveal
layer after layer, the whole complexity of financial price motions.

\pagebreak
\appendix

\section{Proportion of $n$-digit random strings}
\label{annexe1}

In base 2, each digit can be either $0$ or $1$. So, there are at
most 2 different 1-digit binary strings. And more generally, at most
$2^i$ $i$-digit binary strings. So, there are less than
$$ 2+...+2^{h-1}=2^h-2$$
different binary strings that are strictly shorter than $h$.

Then, the proportion of the binary strings, whose length can be
reduced by more than $k$ digits, cannot exceed
$$2^{n-k}-2/2^n<1/2^k.$$

With $k=10$, at most $1/2^{10}=1/1024$ of all $n$-digit binary
strings can be compressed by more than 10 digits. With $k=20$, this
latter proportion cannot exceed $1/1048576$.

\section{ A generating program of the Champernowne's constant}\label{annexe3}

In this appendix, we present the programm - written in
``R''\footnote{\url{http://www.r-project.org/}}- that generates
digits of Champernowne's constant.
\begin{Verbatim}[frame=single]
> c<-0; j<-0 ; d<-0
> for (n in 1:10000)
> {
> a<-n;k<-0
> while (a!=0) {k<-k+1;d[k]<-a%%2;a<-(a-d[k])/2}
> for (h in 1:k) {j<-j+1; c[j]<-d[k-h+1]}
> }
\end{Verbatim}

This programm delivers the first $123631$ digits of Champernowne's
number, while it is written with $132\times8=1056$ digits. In other
terms, this programm realizes a $99.145\%$ compression rate.

\section{Lossless compression
algorithms}\label{annexe4}

Although the intrinsic value of Kolmogrov complexity remains stable
to programming technics, the presence of the constant ``$c$'' in the
invariance theorem (see page \pageref{invariance}) can modify the
compression rate one can obtain on a finite string. Thus, our choice
of compression tools should be as large as possible to estimate the
shortest expression of a given financial series.

From a technical point of view, there are 2 categories of lossless
compression algorithms:
\begin{enumerate}
  \item Entropy coding algorithms reduce file size on exploiting the statistical frequency of
  each symbol. Two technics are often used in this purpose:
  \begin{itemize}
     \item Huffman coding: one of the most traditional text
     compression algorithms. According to this approach, the more
     frequent is a given symbol, the shorter will be its corresponding code in the compressed
     file. Following this principle, Huffman coding is particularly
     powerful on highly repetitive texts.
     \item Dictionary coding: another statistical compression technic used by a big family of
     recent tools, such as Gzip, LZ77/78, LZW. This approach consists to
     construct a one-to-one correspondence between words in the initial text and their code in the compressed file, a so-called ``dictionary''.

     Then, according to the dictionary, each word is ``translated'' into a short
     expression. A positive compression rate can be observed if all words in the
     initial text don't have the same appearance frequency.
     Compared to Huffman coding algorithms, dictionary used in this approach can be modified during the compression procedure. Therefore, dictionary coding algorithms exploit local properties with more efficiency.
     \end{itemize}
  \item Context-based predicting algorithms compress data by forecasting future terms with historical observations. With the development of artificial intelligence, this context-based
approach become more and more performant on self-dependent data. A big
number of technics can be used for data prediction, such as
Prediction by Partial Marching (PPM),
  Dynamic Markov compression (DMC).

PAQ\footnote{Maintained by Matt Mahoney on the website
\emph{http://cs.fit.edu/~mmahoney/compression/}} is one of the most
remarkable family of context-based compression algorithms. These algorithms are reputed by their exceptional compression rates. On
choosing the PAQ version, one can attach more or less importance to
the speed of a compression procedure.

PAQ8o8 is a recent version of PAQ which maximizes the compression rate
at the expense of speed and memory. It's a predicting algorithm which, on analyzing the first
   $t$ digits of a finite string, predicts the appearance probability of each possible symbol at the next digit. The
   $(n+1)th$ digit is coded according to these conditional probabilities.

   In this paper, we tested the performance of the 3 above-cited compression technics on simulated data as well as on Dow Jones daily
   returns, and reported the best compression rate obtained by each
   category of tools. PAQ8o8 delivers by far the best compression rate
   on all discretized series.
\end{enumerate}

\section{Additional figures}\label{annexe5}

  \begin{figure}[h!]
                   \begin{center}
                     \includegraphics[scale = 0.2]{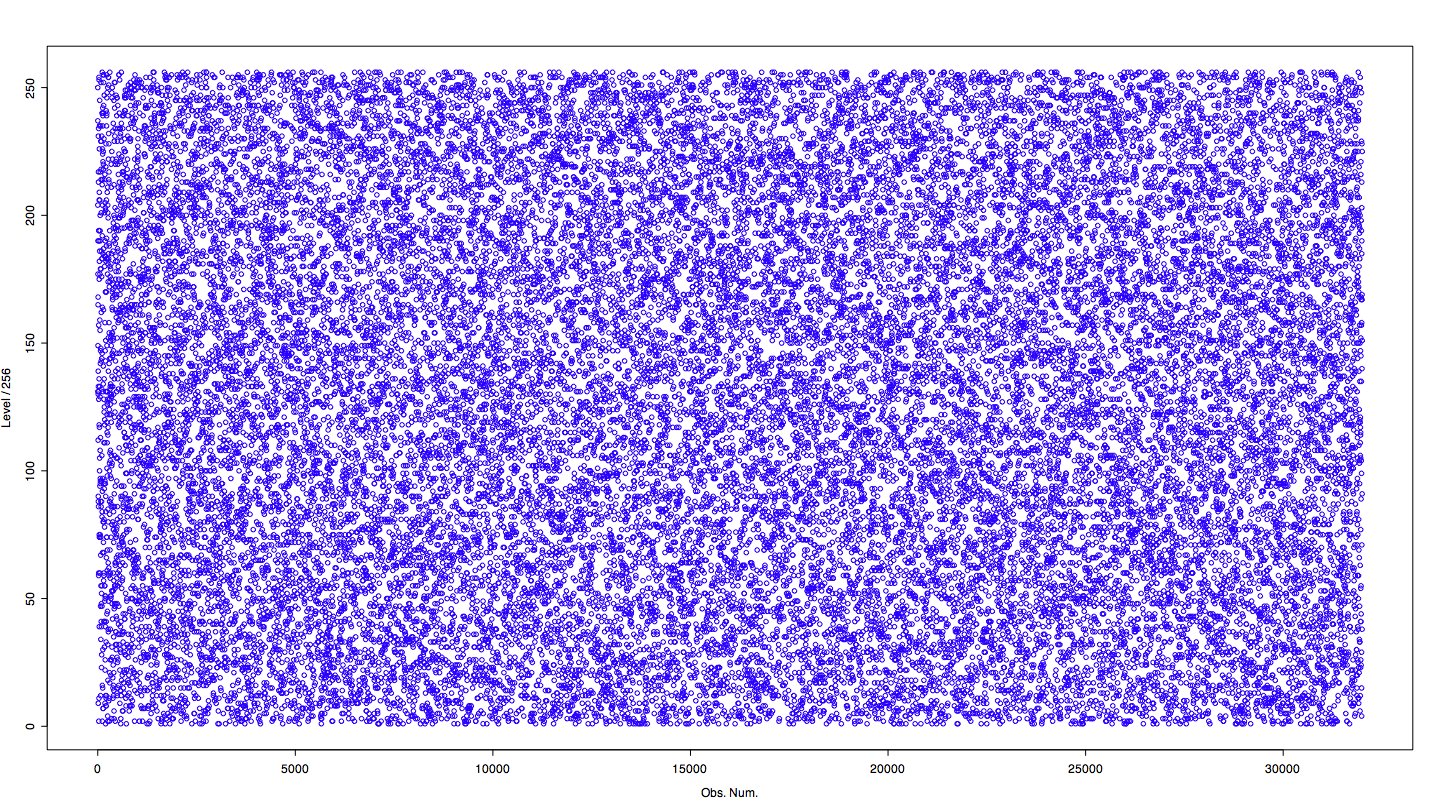}
                   \end{center}
                   \caption{Uniformly discretized returns in section \ref{discret}.}
                   \label{experience2_2}
                 \end{figure}

                     \begin{figure}[h!]
                   \begin{center}
                     \includegraphics[scale = 0.5]{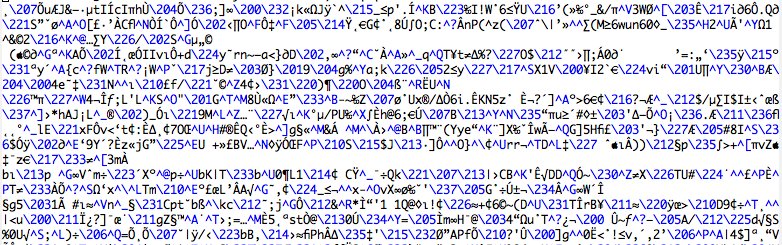}
                   \end{center}
                   \caption{Discretized returns expressed in ascii code}
                  \label{eperience2_3}
                 \end{figure}

\pagebreak
\bibliographystyle{econometrica}
\bibliography{biblioKC2}

\end{document}